\documentstyle[prl,aps,twocolumn]{revtex}
\input{epsf}
\begin{document}

\title{Fermi surface of the colossal magnetoresistance perovskite
La$_{0.7}$Sr$_{0.3}$MnO$_{3}$}

\author{E.A. Livesay$^{1}$, R.N. West$^{1}$, S.B. Dugdale$^{2,3}$,
G. Santi$^{3}$ and T. Jarlborg$^{3}$}

\address{$^{1}$ Physics Department, University of Texas at Arlington,
Arlington,P.O. Box 19059, Texas, TX 76019}

\address{$^{2}$ H.H. Wills Physics Laboratory, University of Bristol, Tyndall 
Avenue, Bristol BS8 1TL, United Kingdom}

\address{$^{3}$ D\'epartement de Physique de la Mati\`ere Condens\'ee,
Universit\'e de Gen\`eve, 24 quai Ernest Ansermet, CH-1211 Gen\`eve 4,
Switzerland} 

\date{\today}
\maketitle
\begin{abstract}
Materials that exhibit colossal magnetoresistance (CMR) are currently the
focus of an intense research effort, driven by the technological
applications that their sensitivity lends them to. Using the angular
correlation of photons from electron--positron annihilation, we present a
first glimpse of the Fermi surface of a material that exhibits CMR,
supported by ``virtual crystal'' electronic structure calculations. 
The Fermi surface is shown to be sufficiently cubic in nature that it is
likely to support nesting.
\end{abstract}


Since the recent discovery of the phenomenon of colossal magnetoresistance
(CMR), research efforts have been intense
\cite{helmholt:93,jin:94,ramirez:97}. The reason is the number of
potentially important applications for CMR materials in magnetic memory
systems, magnetic read heads and in other magnetic sensors
\cite{venkatesan:98}. Experimental studies of CMR materials have
concentrated on the manganite perovskites \cite{ramirez:97},
T$_{1-x}$D$_{x}$MnO$_{3}$, where T is a trivalent lanthanide cation, and D
is a divalent (e.g. alkaline earth) cation. As implied by the Jahn-Teller
distortions manifest in the undoped parent compounds (e.g. LaMnO$_{3}$),
these are systems where there is strong coupling between the electronic and
lattice degrees of freedom \cite{jahnteller}.  That, and the multiplicity
of crystallographic and magnetic phases in the doped crystals, suggests
that the ground states of these systems depend upon a subtle interplay
between their microscopic electrical, magnetic and lattice properties and
excitations \cite{millis:95,singh:98}.  Calculations of electronic band
structures have been reported \cite{singh:98,youn:98}, but experimental
evidence concerning those band structures and their associated electronic
spectra remains scarce.

In particular, a knowledge of the Fermi surface (FS) is vital for an
understanding of transport properties. The half-metallic character (and
therefore the existence of a FS in only one spin) of these materials is
clearly of importance, owing to the absence of a spin-flip scattering
contribution to the resistivity. Moreover, there has been speculation that
one of these sheets has nesting properties \cite{pickett:97}, implying
additional consequences for the transport. In this Letter, we present the
first glimpse of the FS of La$_{0.7}$Sr$_{0.3}$MnO$_{3}$, presented in
conjunction with calculations of the momentum density and band structure.

In such complicated systems, the traditional tools of Fermiology are
precluded, since the disordered nature of the alloy means that the
mean-free-paths are too short. However, the occupied momentum states, and
hence the FS, can be accessed via the momentum distribution using the
2-Dimensional Angular Correlation of electron--positron
Annihilation Radiation (2D-ACAR) technique \cite{west:95}. A 2D-ACAR
measurement yields a 2D projection (integration over one dimension) of the
underlying two-photon momentum density, $\rho(\bbox{p})$ i.e.
\begin{eqnarray}
\rho(\bbox{p}) &=& \sum_{\mbox{occ.}j,\bbox{k}} \vert \int 
d\bbox{r} \sqrt{\gamma(\bbox{r})} \psi_{k,j}(\bbox{r})\psi_{+}(\bbox{r})
\exp (-i\bbox{p}.\bbox{r})\vert ^{2} \nonumber\\
&=& \sum_{j,\bbox{k},\bbox{G}} 
n^{j}(\bbox{k}) \vert C_{\bbox{G},j}(\bbox{k}) \vert ^{2} 
\delta(\bbox{p}-\bbox{k}-\bbox{G}),
\end{eqnarray}
where $\psi_{k,j}(\bbox{r})$ and $\psi_{+}(\bbox{r})$ are the electron and
positron wave functions, respectively, and $n^{j}(\bbox{k})$ is the
electron occupation density in ${\bf k}$-space in the $j^{\mbox{th}}$ band,
and $\gamma(\bbox{r})$ is the so-called enhancement factor which takes
account of electron--positron correlations (and would be unity in the
independent particle model) \cite{jarlborg:87}. The
$C_{\bbox{G},j}(\bbox{k})$ are the Fourier coefficients of the interacting
electron-positron wave function product and the delta function expresses
the conservation of crystal momentum. $\rho(\bbox{p})$ contains information
about the occupied electron states and their momentum, $\bbox{p} = \hbar
(\bbox{k +G})$, and the FS is reflected in the discontinuity in this
occupancy at the Fermi momentum, $\bbox{p_{F}} = \hbar (\bbox{k_{F} +
G})$. If the effects of the positron wave function are small, such that the
$\sum_{\bbox{G}}C_{\bbox{G},j}(\bbox{k})$ are almost independent of ${\bf k}$, a
projection of the ${\bf k}$-space occupation density can be obtained by
folding back $\rho(\bbox{p})$, or its projections, into the first Brillouin
zone (BZ) in accordance with the Lock--Crisp--West (LCW) prescription
\cite{lock:73}.

It is also possible to exploit the partial ($\sim$ 10\%) polarization of
the positron parallel to the emission direction from the $^{22}$Na source,
arising from the parity-non-conserving $\beta$-decay. This positron
polarization persists throughout the thermalization process
\cite{page:59,blank:88}. If the polarity of the $\sim$1T magnetic field
that focuses the positrons onto the sample is switched (reversing the
direction of magnetization in the sample), then the annihilation
probabilities with the electrons in the majority/minority bands are
modified in such a way as to enable the measurement of the spin-density in
momentum space. During the experiments, the field was reversed every two
days in order to minimize any systematic effects owing to the decay of the
$^{22}$Na source ($T_{1/2} \sim 2.6$ years).

The La$_{0.7}$Sr$_{0.3}$MnO$_{3}$ sample was cut with a diamond saw from a
cylinder grown, using the floating-zone technique, by the group of Tokura
\cite{urushibara:95}.  The specimen was then mechanically
polished. Although the structure of La$_{0.7}$Sr$_{0.3}$MnO$_{3}$ is
distorted from the ideal perovskite, the theoretical calculations and the
treatment of the experimental data were made, for simplicity, under the
assumption of an undistorted cubic perovskite. The sample was aligned using
Laue x-ray back-reflection to ensure that the projection direction would be
down the crystalline [001] axis.  The experimental spectra were measured on
the UTA 2D-ACAR spectrometer \cite{west:81} at a temperature of
$\sim$30K. The FWHM of the total experimental resolution, which is well
described by a Gaussian, was of the order of 7\% of the size of the
Brillouin zone (BZ).  A total of $\sim$400 million counts were
accumulated. After verification that the spectrum exhibited the appropriate
symmetry, that symmetry was then forced upon it by folding, thereby
increasing the effective number of counts to more than three billion.

The spin-dependent momentum densities were calculated using the linearized
muffin-tin orbital (LMTO) method within the atomic sphere approximation
(ASA), including combined-correction terms \cite{andersen:75}. The
exchange-correlation part of the potential was described in separate
calculations in both the local spin-density (LSDA) and generalized-gradient
approximations (GGA, \cite{perdew:85}), but there was no significant
difference in the results obtained. Consequently, those presented here use
the LSDA.  The self-consistent band-structure was calculated at 364
$k$-points in the irreducible 1/48$^{th}$ part of the BZ using a basis set
of $s$, $p$, and $d$ functions. The lattice parameter was set to a value of
3.89\AA, obtained from a linear interpolation of the experimental values
\cite{urushibara:95}. The electronic wave functions were then used to
generate the electron momentum densities for the up- and down-spin bands
separately. In the construction of the momentum density, 2095 reciprocal
lattice vectors were used. A full description of the technique is given in
the papers of Andersen \cite{andersen:75} and Singh and Jarlborg
\cite{singh:85}. To simulate the Sr$^{2+}$ doping, a ``virtual crystal''
approach was employed, whereby the self-consistency was realised in a cell
with the La$^{3+}$ ions replaced by virtual ``2.7+'' ions (i.e. with 2.7
protons). While being a relatively vulgar way of describing the doping, our
calculations and those of others (for example, see
\cite{pickett:97,youn:98}), have shown it to explain the systematic trends
in the electronic structure.  The calculated momentum distributions were
numerically integrated (along the [001] direction) for direct
comparison with the experiment.

\begin{figure}
\epsfxsize=230pt
\epsffile{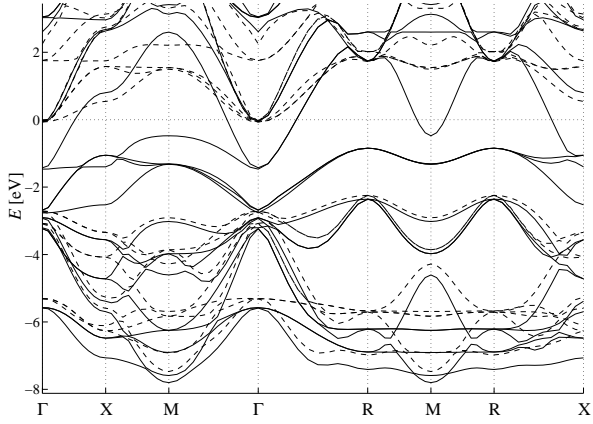}
\caption{Spin-polarized band structure of
La$_{0.7}$Sr$_{0.3}$MnO$_{3}$. The majority bands are shown as solid lines,
and the minority as dashed lines. Note that the Fermi energy lies just above a
gap in the minority bands.} 
\label{bands}
\end{figure}

In Fig.~\ref{bands}, the band structure along the usual high-symmetry
directions is plotted. The bands are similar to those calculated by Pickett
and Singh \cite{pickett:97} and Youn and Min \cite{youn:98}.  The Fermi
level lies just above a gap in the minority density-of-states, indicating
that the system is very close to being half-metallic. At the Fermi level,
the bands have mainly Mn $d$ character (these are the $e_{g}$ bands~; the
$t_{2g}$ bands are the more localised set of bands, lying a couple of eV
below the Fermi energy). However, it can be seen that there are two small
minority electron sheets~; a tiny downward shift in the Fermi energy would
make these disappear, the system becoming half-metallic.

\begin{figure}
\epsfxsize=230pt
\epsffile{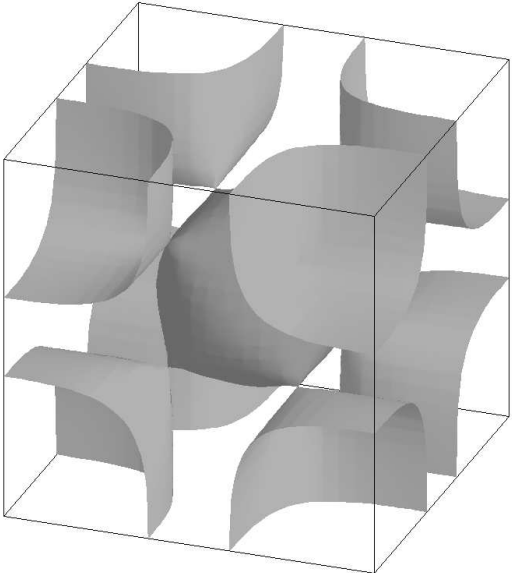}
\caption{Two sheets of the Fermi surface of
La$_{0.7}$Sr$_{0.3}$MnO$_{3}$. Hole cuboids at the $R$ points, coined
``woolsacks''. Electron spheroid centered at the $\Gamma$ point.} 
\label{FS}
\end{figure}

Fig.~\ref{FS} shows the two principal majority FS sheets coming from the
LMTO calculation. These comprise hole cuboids at the $R$ points (coined
``woolsacks'') that touch an electron spheroid centered at the $\Gamma$
point along the $<$111$>$ directions. These woolsacks, with their
relatively flat faces, present the possibility of carrier ``skipping'' along
the parallel sections of surface (with the initial and final velocities
remaining parallel), in addition to supporting a FS capable of nesting
\cite{pickett:97}. Both of these have implications for the transport
properties owing to the concentration of phase-space for the scattering
along directions parallel to the cube edges, as noted by Pickett {\it et al.} for
the flat, quasi-one-dimensional parts of the FS of the cuprates
\cite{pickett:90}. Clearly, the degree of flatness of the sides of these
``woolsacks'' will determine the strength of these nesting and skipping
tendancies~; we will return to this later when we discuss the experimental
results.

\begin{figure}
\epsfxsize=230pt
\epsffile{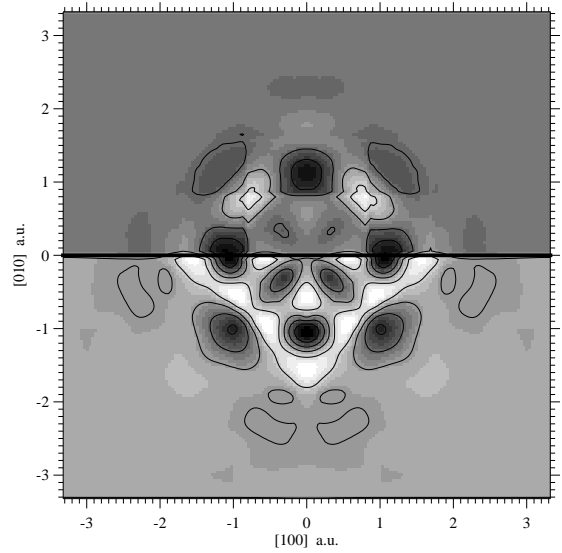}
\caption{Radial anisotropy of the [001]-projected momentum density, coming
from the experiment (top) and LMTO calculation (bottom).}
\label{anisotropy}
\end{figure}

The radial anisotropy of the data and LMTO calculation is plotted in
Fig.~\ref{anisotropy}. This is constructed by subtracting the radial
average of the spectrum from the spectrum itself. The agreement between the
experiment and theory is excellent.  So as not to obscure any details, the
theory has not been convoluted with the experimental resolution
function. In complex, multi-band systems, it is often the case that the
anisotropy is dominated by the fully-occupied bands, rather than the
valence bands, and as such does not reflect the topology of the underlying
FS. An LMTO calculation of the anisotropy from just the oxygen sublattice
shows that this provides the dominant contribution to the total anisotropy
i.e. as the states at the Fermi level are predominantly of Mn
$d$ character, the total anisotropy is not going to reflect the FS.

\begin{figure}
\epsfxsize=230pt
\epsffile{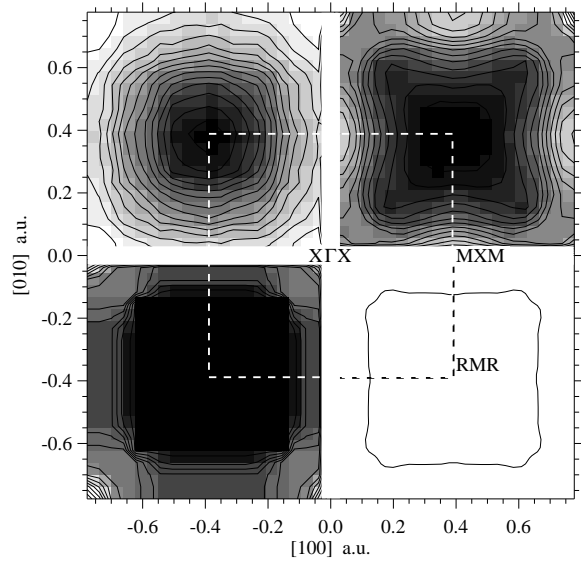}
\caption{{\bf Top left} : Occupancy projected along [001], coming from the
LCW-folded experimental spectra. {\bf Top right} : LCW of calculated
electron-positron momentum density. {\bf Bottom left} : Calculated electron
occupancy. {\bf Bottom right} : The ``woolsacks'' FS, extracted from the
experimental data by the zero-crossing procedure outlined in the main text.
The Brillouin zone is marked by the dotted line, and the labels indicate
the projected symmetry points.}
\label{LCW}
\end{figure}

The LCW-folded data, together with the calculated electron, and
electron--positron occupancies (projected down [001]) are shown in
Fig.~\ref{LCW} as gray-scale images, the white areas being the areas with
the highest occupancy.  Aside from some limited evidence for electron
pockets at $\Gamma$, the dominant features are the woolsack hole
pockets (see Fig. \ref{FS}) centered on the $R$ points of the BZ. These can
be clearly identified in both the experimental and theoretical
distributions, but owing to the smearing of the experimental resolution
(and to some extent, perturbations introduced by the positron wave
function), their size and shape is not clear. Additionally, the finite
sampling in the calculations may slightly degrade their resolution.

Recent advances in the analysis methods for 2D-ACAR data have opened up the
possibility for the extraction of extremely reliable and accurate
information about the shape and size of the Fermi surface, even though the
spectra are smeared with the experimental resolution and influenced by
positron wave function effects. In most cases, a full 3D reconstruction is
needed \cite{west:95}, and hence it is necessary to measure a series of
projections along different directions. However, if the FS topology is
relatively simple, as is the case here, this information can be gleaned
from just one projection. This is because the hole cuboids of interest
project onto themselves, and are not obscured by any other FS feature.
Previously, it was proposed by Dugdale {\it et al.} \cite{dugdale:94,dugdale:96}
that it is possible to define the FS by a zero-crossing contour in a
filtered distribution. The filtering method used there was based on Maximum
Entropy, but band-pass filters were being simultaneously employed for a
similar purpose \cite{obrien:95}. More recently, these techniques were used
to reveal the FS of yttrium, and extract quantitative information about a
particular sheet \cite{dugdale:97}. In the bottom right of Fig.\ref{LCW},
this zero-crossing contour is plotted, and it is evident that it is very
square in shape.  (A similar procedure applied to the electron--positron
calculation gave very similar results).  This supports the idea that this
sheet of FS does indeed have the nesting and skipping properties referred
to earlier. In addition, the size of the cube has been determined as $0.65
\pm 0.02 \times (2\pi/a)$. This compares reasonably with the value of $0.78
\times (2\pi/a)$ taken from the LMTO theory (and that of $0.81 \times
(2\pi/a)$, found by Pickett and Singh \cite{pickett:97}).  The
$\Gamma$-centered electron surface does not appear to be defined by this
zero-contour. This could be because its shape is more spherical than the
flat-sided hole cubes and hence does not generate as sharp an edge in the
projection, thus not being as ``enhanced'' by the band-pass filter.

\begin{figure}
\epsfxsize=230pt
\epsffile{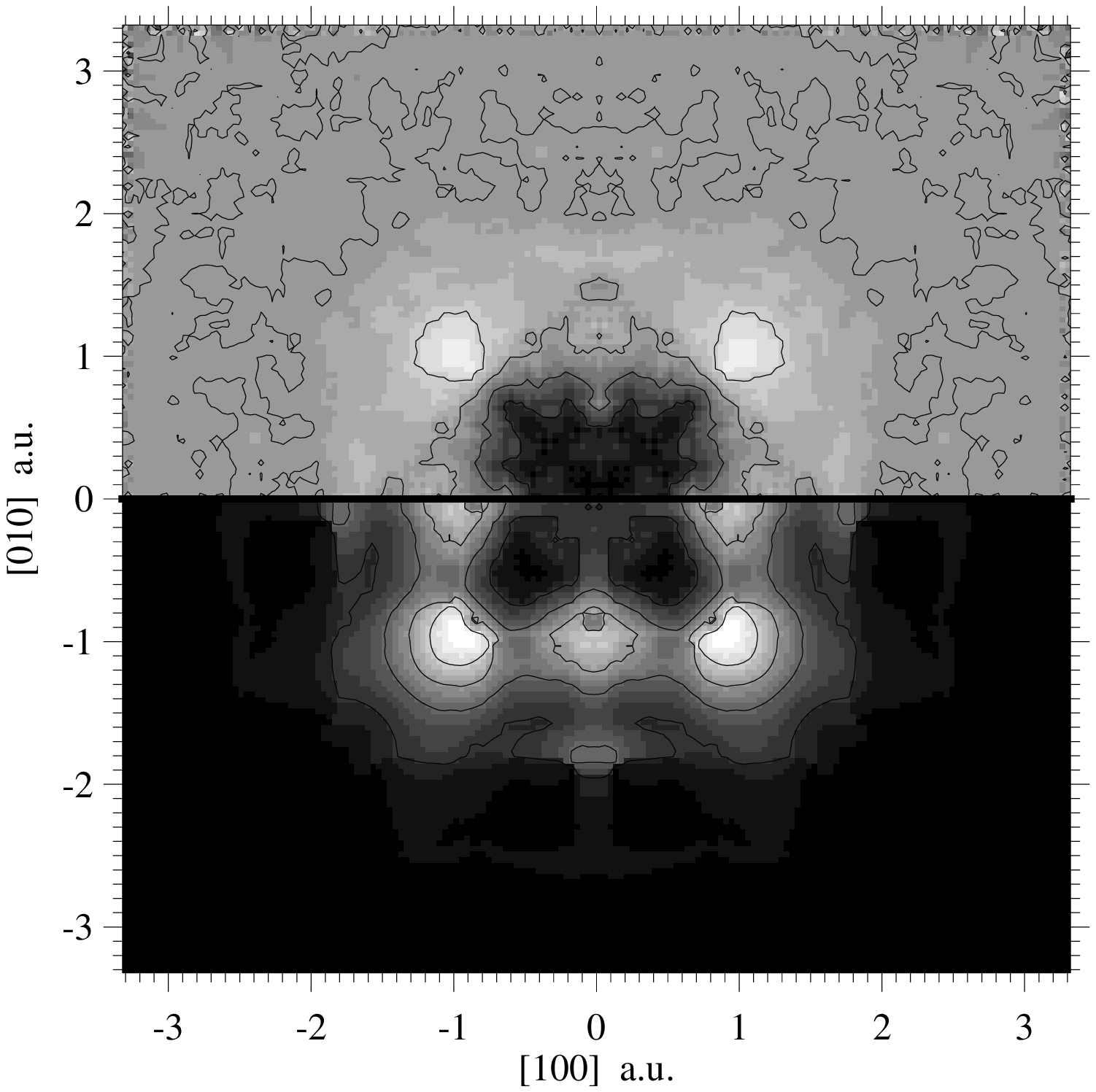}
\caption{The spin density in momentum space as seen by the positron,
integrated along the [001] direction from the experiment (top) and LMTO
calcuation (bottom). White areas indicate positive spin-polarisation.}
\label{spin}
\end{figure}

Finally, in Fig.~\ref{spin} we present the preliminary results of the
experimentally measured spin density, constructed by subtracting the
spectra measured with the sample magnetizations in opposite directions,
along with the theoretically calculated counterpart. The agreement between
the experiment and calculation is excellent. That the excess spin is
predominantly in the larger momentum regions is consistent with its origin
in the $d$ bands.  A more detailed examination of these spin distributions
is underway.
 
In conclusion, we have measured the Fermi surface topology of
La$_{0.7}$Sr$_{0.3}$MnO$_{3}$ using the 2D-ACAR technique.  Our results
agree well with the findings of our own band structure calculation and with
those of others \cite{pickett:97,youn:98}. In particular, using recently
implemented methods for determining the FS, we have shown the existence of
a cuboid hole FS sheet, centered on the $R$ point of the Brillouin
zone. The cuboid has been shown to have a side of $0.66 \pm 0.02 \times
(2\pi/a)$, and to be sufficiently flat for the carrier ``skipping'' and
nesting described by Pickett and Singh \cite{pickett:97} to be extremely
favorable.

We are particularly indebted to Prof. Tokura (JRCAT, Japan) for providing
the sample. We would also like to acknowledge support from the Texas
Advanced Research Program (Grant No. 003656-142). One of us (SBD) would
also like to thank the Lloyd's of London Tercentenary Foundation for the
provision of his fellowship, and the EPSRC (UK) for other support.

\begin{thebibliography}{99}


\bibitem{helmholt:93}
R.~M. von Helholt {\it et al.}, Phys. Rev. Lett. {\bf 71}, 2331 (1993).

\bibitem{jin:94}
S. Jin {\it et al.}, Science {\bf 264}, 413 (1994).

\bibitem{ramirez:97}
A.~P. Ramirez, J. Phys.: Condens. Matter {\bf 9}, 8171 (1997).

\bibitem{venkatesan:98}
T. Venkatesan {\it et al.}, Phil. Trans. Roy. Soc. Lond. A {\bf 356}, 1661 (1998).

\bibitem{jahnteller}
S. Satpathy {\it et al.}, Phys. Rev. Lett. {\bf 76}, 960 (1996)~; Solovyev et
al., Phys. Rev. Lett. {\bf 76}, 4825 (1996).

\bibitem{millis:95}
A.~J. Millis, P.~B. Littlewood and B.~I. Shraiman, Phys. Rev. Lett. {\bf
75}, 5144 (1995)~; A.~J. Millis, B.~I. Shraiman and R. Mueller,
Phys. Rev. Lett. {\bf 77}, 75 (1996).

\bibitem{singh:98}
D.~J. Singh and W.~E. Pickett,  Phys. Rev. B {\bf 57}, 88 (1998).

\bibitem{youn:98}
S.~J. Youn and B.~I. Min, J. Korean Physical Society {\bf 32}, 576 (1998).

\bibitem{pickett:97}
W.~E. Pickett and  D.~J. Singh, Phys. Rev. B {\bf 55}, R8642 (1997).

\bibitem{west:95}
R.~N. West,  in {\em Proceedings of the International School of Physics
  $<<$Enrico Fermi$>>$ --- Positron Spectroscopy of Solids}, edited by A.
  Dupasquier and A.~P.~Mills jr. (IOS Press, Amsterdam, 1995).

\bibitem{jarlborg:87}
T. Jarlborg and A.~K. Singh, Phys. Rev. B {\bf 36}, 4660 (1987).

\bibitem{lock:73}  
D.~G. Lock, V.~H.~C. Crisp, and R.~N. West, J. Phys. F {\bf 3},  561  (1973).

\bibitem{page:59}
L.~A. Page, Rev. Mod. Phys. {\bf 31}, 759 (1959).

\bibitem{blank:88}
R. Blank, L. Schimmel and A. Seeger, in {\em Positron Annihihilation}
ed. L. Dorikens {\it et al.}, p.~278 (World Scientific, Singapore, 1988).

\bibitem{urushibara:95}  
A. Urushibara {\it et al.}, Phys. Rev. B {\bf 51}, 14103 (1995).

\bibitem{west:81}
R.~N. West, J. Mayers and P.A. Walters, J. Phys. E {\bf 14}, 478 (1981).

\bibitem{andersen:75}
O.~K. Andersen, Phys. Rev. B {\bf 12}, 3060 (1975)~;
T. Jarlborg and G. Arbman, J. Phys. F {\bf 7}, 1635 (1977).

\bibitem{perdew:85}
J.~P. Perdew, Phys. Rev. Lett. {\bf 55}, 1665 (1985) .

\bibitem{singh:85}
A.~K. Singh and T. Jarlborg,  J. Phys. F {\bf 15}, 727 (1985).


\bibitem{pickett:90}
W.~E. Pickett, H. Krakauer and R.~E. Cohen,  Physica B {\bf 165 \& 166},
1057 (1990)~; H. Krakauer, W.~E. Pickett and R.~E. Cohen,  Phys. Rev. B
{\bf 47}, 1002 (1993).

\bibitem{dugdale:94}
S.~B. Dugdale {\it et al.}, J. Phys.:Condens. Matter {\bf 6}, L435 (1994). 

\bibitem{dugdale:96}
S.~B. Dugdale, Ph.D. thesis, University of Bristol, UK (unpublished) (1996). 

\bibitem{obrien:95}
K.~M. O'Brien, M.~Z. Brand, S. Rayner and R.~N.West, J. Phys.: Condens. Matter
{\bf 7}, 925 (1995).

\bibitem{dugdale:97}
S.~B. Dugdale {\it et al.}, Phy. Rev. Lett. {\bf 79}, 941 (1997).

\end{thebibliography}
\end{document}